\documentclass[cmfonts]{witpress}
\usepackage[latin1]{inputenc}
\usepackage[english,brazil]{babel}
\usepackage{graphicx}
\usepackage{table}
\usepackage[pdftex]{hyperref}

\title{A water level relationship between consecutive gauge stations along  Solimões/Amazonas main channel: a wavelet approach}

\author{\large{R. D. Somoza ; E. S. Pereira ;  E. M. L. Novo;  C. D. Rennó}}

\address{Department of Remote Sensing,\\ National Institute of Space Research,\\ Brazil}

\begin{document}
\pretolerance=10000
\maketitle
\textbf{{\large Abstract}}

Gauge stations are distributed along the Solimões/Amazonas main channel to monitor water level changes along the time. Those measurements  help quantifying both the water  movement and its variability from one gauge station to the next downstream. The objective of this study is to detect changes into the water level  relationship  between consecutive gauge stations  along the  Solimões/Amazonas main channel, since 1980. To carry out the analyses, data spanning from 1980 to 2010 from three consecutive gauges (Tefe, Manaus and Óbidos) were used to compute standardized daily anomalies. In particular for infra-annual periods was possible to detect changes for the water level variability along the Solimões/Amazonas main channel, by applying the Morlet Wavelet Transformation (WT) and  Wavelet Cross Coherence (WCC) methods. It was possible to quantify the waves amplitude for the WT infra-annual scaled-period and were quite similar to the three gauges station denoting that the water level variability are related to a same hydrological forcing functions. Changes in the WCC was detected for the Manaus-Óbidos river stretch and this characteristic might be associated with land cover changes in the floodplains.The next steps of this research, will be to test this hypotheses by integrating land cover changes into the floodplain with hydrological modelling simulations throughout the time-series.


\section{Introduction}

For any large river basin, the annual hydrological process  are very important in terms  of geomorphology, biochemistry and sediment transport, as well as the biota and biodiversity maintenance into the floodplain \cite{Junk1997,ALCANTARAetal2010}. Surface fresh waters are also important for the management of water resources use by people. One of the biggest challenges in the 21st century will be managing water availability and related global health and hazard prevention aspects \cite{Papaetal2008}.

Nowadays, the knowledge of space and time variations of continental waters  relies on \textit{in situ} gauge measurements and hydrological modelling. \textit{In situ} gauge stations measurements help quantifying both the water  movement and its variability from one gauge station to the next downstream. Changes in both, water movement and its variability along the time may be related to environmental changes in the river floodplain. 

Remote sensing surveys along the Solimões/Amazonas main channel have shown large differences in floodplain land cover upstream and downstream Manaus \cite{Hessetal2003}. In the last 30 years, almost half of the flood forest have been removed between Parintins and Almeirim \cite{Renoetal2011Viv}. From the hydrodynamic point of view, the roughness on this deforested areas move from a soft to a stiff blade, increasing both the friction coefficient and the superficial water velocity.

On this context, were selected the Manaus gauge station and two others representing upstream and downstream Manaus to explore the fluviometric measure data throughout the time series. The main objective of this study is to detect changes in the water level  relationship  between consecutive gauge stations  along the  Solimões/Amazonas main channel, since 1980, by applying a wavelet approach.

\subsection{Data and methodology}
\subsection{Study Area}

\begin{figure}[h!]
 \begin{center}
  \includegraphics[width=100mm, height=6cm]{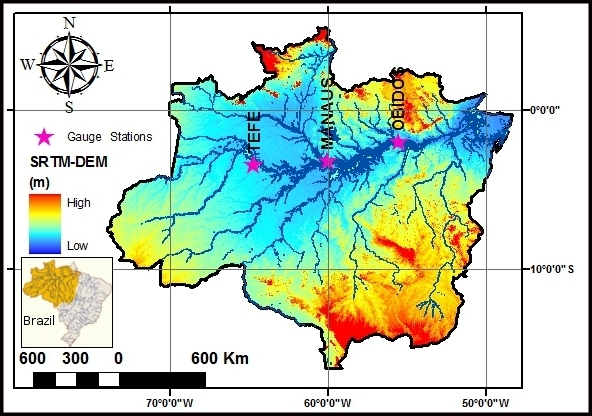}\\
   \caption[]{Study Area: Amazon River Basin}
    \label{fig:1}
 \end{center}
\end{figure}

The Amazon Basin is considered the largest hydrographic river basin in the world with $~6.1$ million km$^{2}$, extending from the Andes to the Atlantic Ocean (Figure \ref{fig:1}). In tropical areas, hydrographs of large river systems such as the Amazon River often have a peak annual flood, resulting from seasonal changes in precipitation \cite{Birkett1998}. 

The large range of annual variation in water level and inundated area (rivers and floodplains) in the Amazon basin is responsible for large seasonal changes in water mass \cite{ALMEIDA2012}. The distribution of  floodplain environments along the Solimões/Amazon River is strongly influenced  by the rise and fall of its water level. In rivers with low slope gradients such as the Amazon, the input peak is usually higher than the output peak, resulting in a time lag in the hydrographs \cite{KazezyilmazAlhaneMedina2007}. Due to the Amazon basin size, this time lag, can be larger than on month after the rainy season, in the Central Amazon near Manaus \cite{Junk1997}. 

\subsection{Gauges Stations}

Three gauge station distributed form upstream to downstream were selected for this study: Tefe, Manaus and  Óbidos(Figure \ref{fig:1}).These stations are assumed to represent water level variation in the Amazon/Solimões main channel even when they are  not exactly located in the main stem such as the case of Manaus gauge station. Those stations   have an historical series of fluviometric records larger than 30 years (Table \ref{tabela:1}). All data was obtained freely from the Brazilian Hydrological Information System website (HIDROWEB, \url{http : //hidroweb.ana.gov.br/})

\begin{table*}[h]
	    \caption{Fluviometric gauge station description. Water level-period is refering to the 1$^{st}$ Measurement when the time series begin.}
	    	\label{tabela:1}
	    	\begin{center}
\begin{tabular}{|c|c|c|c|}
Gauge Station & Latitude  & Longitude &  Water level-period  \\\hline

TEFÉ & S $3^{o} 22' 33.6"$ & W $64^{o} 39' 18.0"$  & Aug/1982 \\\hline

MANAUS & S $3^{o} 8' 13.20"$ & W $60^{o} 1' 37.20"$  & Sep/1902 \\\hline

ÓBIDOS & S $1^{o} 55' 8.40"$ & W $55^{o} 30' 46.80"$ & Dec/1927 \\\hline

\end{tabular} 
\end{center}
\end{table*}

Since 1980, all available data of water level daily records from Manaus and Óbidos were used to carry out the analysis, while from Tefe the data was available since late 1982. Each data series was preprocessed in order to remove the outliers values and complete the series by using linear interpolation for gaps smaller than three days. Gaps larger than three days remain in the series and were not considered in the analyses. The Tefe data series had gaps  from 1 week to 6 month on the following years: 1983, 1987, 1988, 1990, 1991, 1994 and 1995. There were not gaps in the Manaus and Óbidos data series. 

\subsection{Computing Method}

To carry out the analyses, data spanning from 1980 to 2010,  described in the Table \ref{tabela:1},  were used to compute the water level standardized anomalies \cite{Nobreetal2006}. First, the Morlet wavelet transformation (WT) was applied to each series and being computed 5 \% of significance against the red noise model and the influence cone  \cite{TorrenceCompo1998}. It was computed the Wavelet Cross Coherence (WCC) power spectrum of the  water level standardized daily anomalies relations between consecutive Gauge stations: Tefe-Manaus,  Manaus-Óbidos \cite{Grinsted04,TorrenceCompo1998}.  In order to validate the Wavelet Coherence, a Monte Carlo simulation was applied to obtain a 95 \% confidence level  for the WC  against the red noise model as described in \cite{Grinsted04}.

To carry out this study, a program written in python language was developed to help integrate all data and results into a relational data base (Sqlite). The connection with the data base and the program was made using the Database Abstract Layer (DAL) from the framework web2py \cite{pierro09}. This DAL organization provides advantages for consulting, manipulation and interoperability  by date.

For the wavelet analysis a Python programming interface among routines originally written in Matlab by \cite{Grinsted04} was developed. These routines was accessed in python, by octave(\url{http://www.gnu.org/software/octave/}) language, using oct2py package (More details in: \url{https://pypi.python.org/pypi/oct2py}). However, some functionality of Matlab were not available in octave, in this case, a minor modification of the routines from  \cite{Grinsted04}  was done. The plot functions was reimplemented  in pure python using matplotlib package (More details in: \url{https://pypi.python.org/pypi/matplotlib/1.2.0}). The final version of this Python programming interface package,  that it was called as \textbf{Piwavelet},  was released under general GNU license version 3, and, it was located for free access in the web site  \url{http://duducosmos.github.com/PIWavelet/}.

\section{Wavelet Transform}

The Wavelet Transform (WT) has emerged in recent years as a powerful time-frequency analysis and signal coding tool favoured for complex non-stationary signals. It is a tool for characterizing the frequency, the intensity, the time position, and the duration of variations in hydro-meteorological series \cite{Zhanetal2010}. Using WT, time series can be decomposed into time-frequency space, determining both the dominant modes of variability and how these modes vary in time \cite{TorrenceCompo1998}.

\begin{figure}[h!]
 \begin{center}
  \includegraphics[width=130mm]{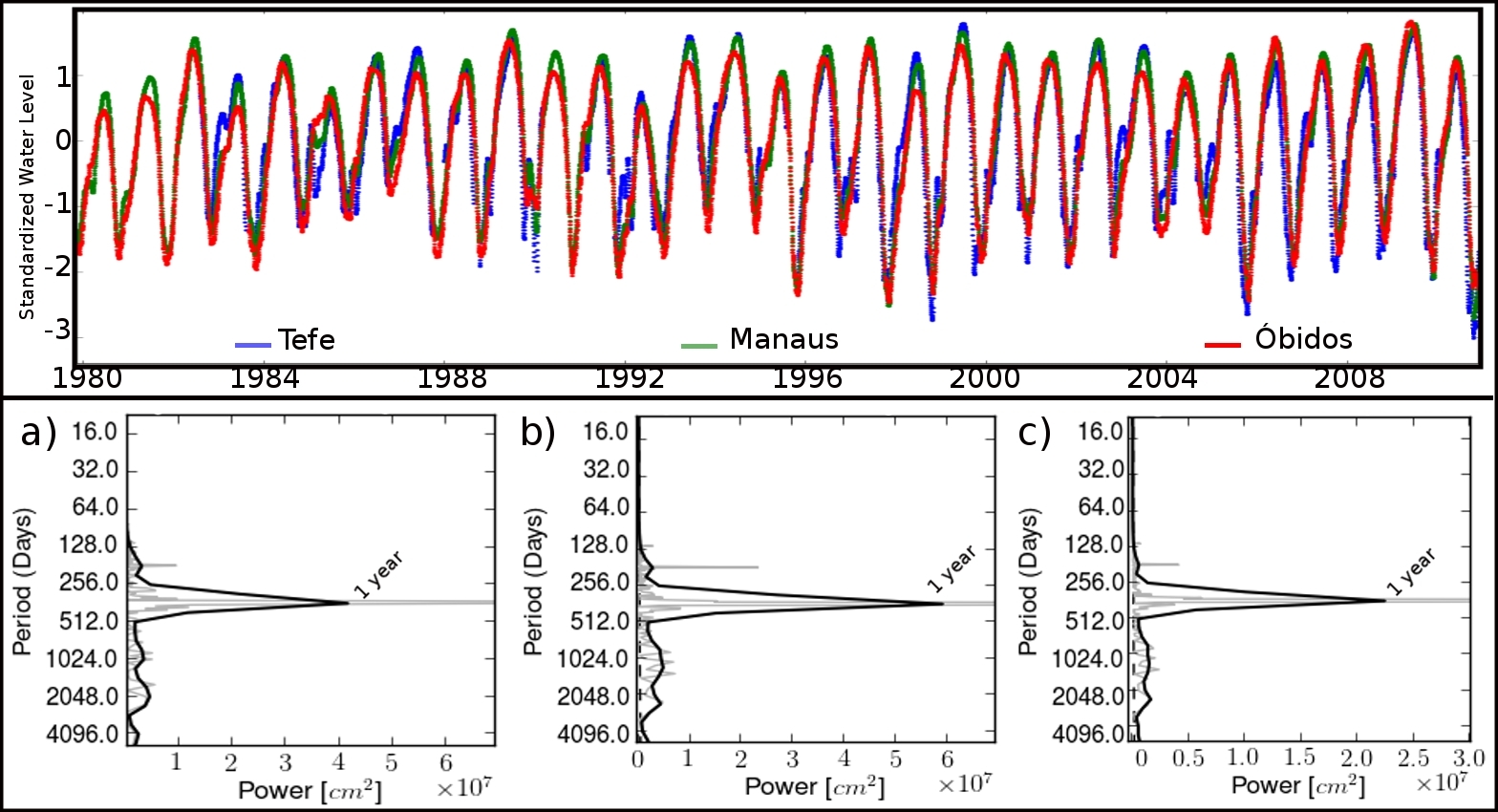}\\
   \caption{Standardized daily  water level on the top, Y-axis are the anomalies values ($cm/cm$) and the X-axis are time (years).  Global wavelet spectrum ($cm^{2}$), for the Gauge stations: (a) Tefe, (b) Manaus, and (c) Óbidos.}
    \label{fig:W1}
 \end{center}
\end{figure}

Figure \ref{fig:W1} depicts the standardized daily  water level  and the global wavelet spectrum (WT integration throughout the time-series) for the tree gauge stations, over the 1980-2010 period. The water level, for all gauges, in consequence of their locations in the tropics,  is characterised by a strong annual cycle and superimposed with a visible inter-annual variability \cite{Marengoetal1998} (Figure \ref{fig:W1} on top). Considering annual and infra annual scales, the Global power spectra shows an strong annual components (Power $\approx 10^{7}~cm^{2}$) for all wavelet signals. This characteristic is masking the other signals energy that might came from higher or lower frequencies.

\begin{figure}[h!]
 \begin{center}
  \includegraphics[width=12cm, height=10cm]{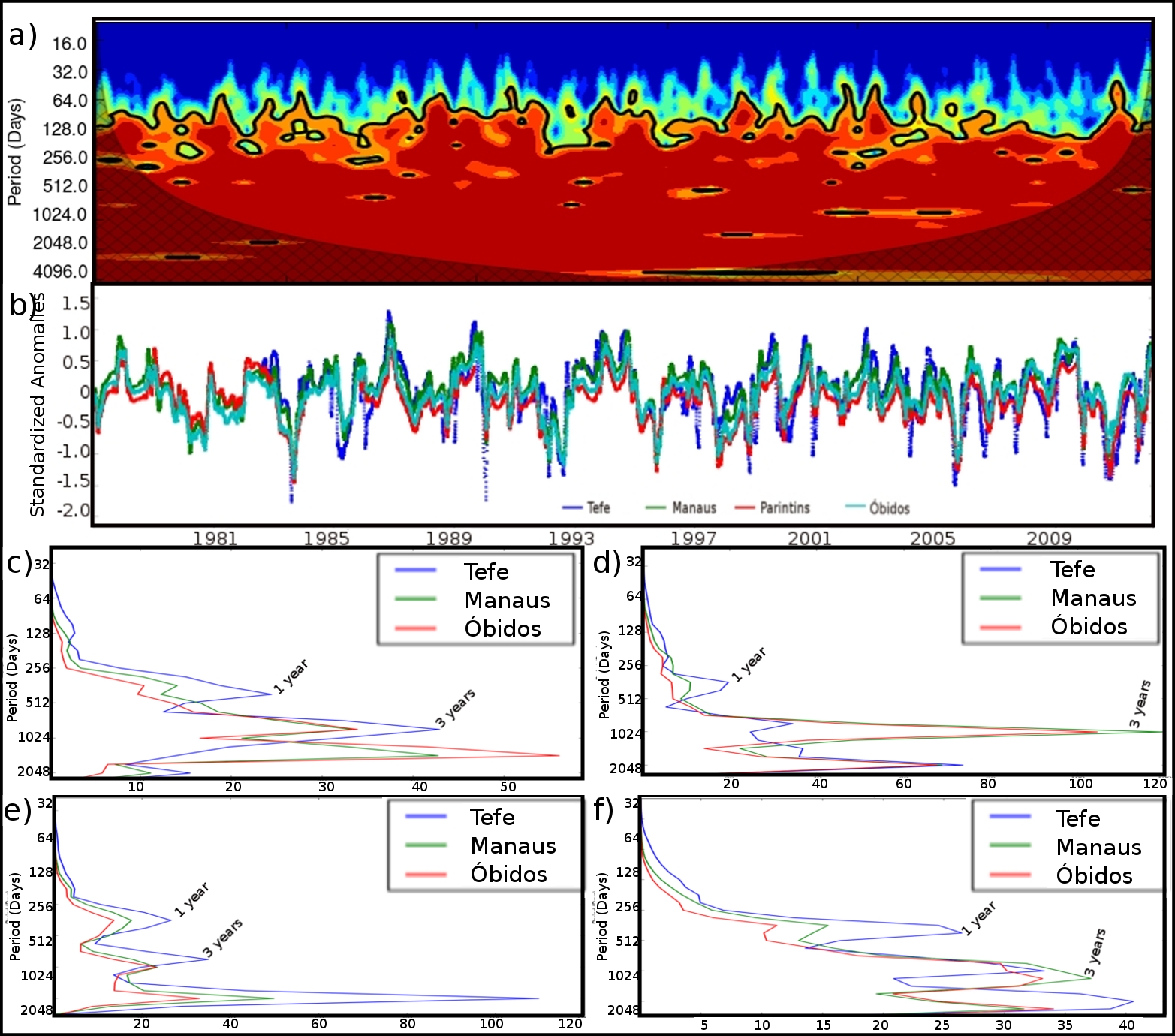}\\
   \caption{ (a) Morlet Wavelet power spectrum for the Manaus Gauge stations. Y-axis is wavelet scale-period (days), and the X-axis are time (years). The U-shaped grid form  shows the cone of influence. The 5\% significance level against red noise model is shown as a thick black contour.   (b) Water level standardized daily anomalies (cm/cm) for Tefe, Manaus and Óbidos gauge stations. The (c), (d) and (e) boxes are the global wavelet power spectrum of these gauge stations for the three decades, 1980s, 1990s, 2000s respectively. The box (f) represents the global power spectrum for 1980-2010 time-series.}
    \label{fig:W2}
 \end{center} 
\end{figure}

Due the annual climatological cycle were very strong, this feature was removed from the historical series by using the daily climatological mean into the standardized anomalies calculi (Figure \ref{fig:W2} b). This way, the global wavelet spectrum (GWPS) was zoomed to make it possible to observe multi-annual scales, which have already been described in literature. For example, Óbidos gauge station has been described  by the wavelet of watershed monthly mean variability and characterizes by the inter-annual 3 year processes coherence with Teleconnection Indexes \cite{Labatetal2005}.

In the Morlet Wavelet power spectrum (Figure \ref{fig:W2} a) it was observed a high variability band between 2-4 month (64-128 days) scale-period. On this band the WT was oscillating between time-periods with high and low synergism emphasizing a high frequency flood pulse. Even removing the annual variability, there was a remaining peak in this period (Figure \ref{fig:W2} f). Also it was  possible to observe the 3-year processes behaviour into the global wavelet spectrum  and another peak on about 6 years  for the all series plotted. These 3 year processes, for example, have been related to the South Oscillation Index, to the Sea Surface Temperature and other  Teleconnection Indexes \cite{Marengoetal1998, Uvoetal1998,  Labatetal2005,Nobreetal2006}.

On the other hand, from the decadal global wavelet power spectrum (Figure \ref{fig:W2}  c, d, e), it was possible to observe changes in the amplitude distributions as a function of the period-scale. In particular, the  three years peak have presented a strong amplitude variation for the last 30 years. In the infra-annual band of 32-128 days minor changes were noted, suggesting  that, from the 1980s to the 2000s decades the associated waves amplitude have decreased, while for  the band 128-256 scale-period was observed an amplitude increase. In the Table \ref{tabela:serie} those changes are quantified by integrating the decadal GWPS in the 32-128  scale-period.
\begin{table*}[h!]
	    \caption{ Integrated decadal global wavelet power spectrum in the 32-128 and 128-256 scale-period.}
	    	\label{tabela:serie}
	    	\begin{center}
\begin{tabular}{|c|c|c|c|}
 Scale-period and decade &  Tefe (cm$^{2}$day)  & Manaus (cm$^{2}$day)  &  Óbidos (cm$^{2}$day) \\ 
\hline 32-128, 1980s  &  141.45 &  48.95  &  28.36  \\ 
\hline 32-128, 2000s  &  114.50 &  39.35  &  23.85 \\ 
\hline 128-256, 1980s &  469.24 &  299.78 &  185.81  \\ 
\hline 128-256, 2000s &  661.66 &  490.78 &  331.53 \\ 
\hline 
\end{tabular} 
\end{center}
\end{table*}

Even the global WT characterization were quite similar to the three gauges station denoting that the water level variability are related to a same hydrological forcing functions, it was possible to quantify the waves amplitude for the infra-annual scaled-period. The more important point in here is the possibility of observing changes for the water level variability along the Solimões/Amazonas main channel, in particular for infra-annual scale,  in the last three decades by applying the WT. A  wavelet cross coherence spectrum will allow to measure the degree of wavelets coherence relation in two consecutive gauge stations as it is going to be presented in the next section. 

\section{Time-scale relationship between two signals}

\begin{figure}[h]
  \begin{center}
   \includegraphics[width=130mm]{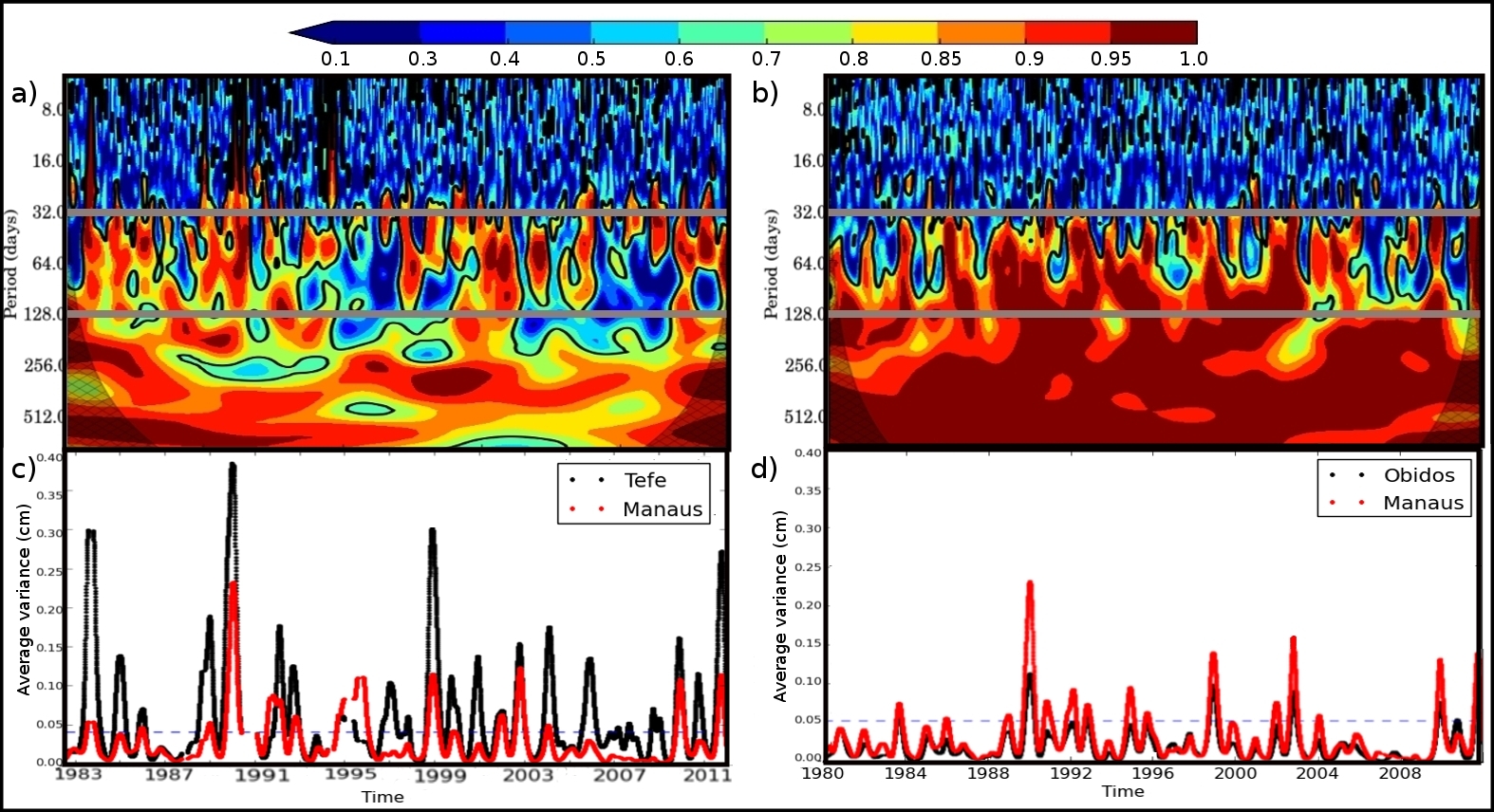}\\
    \caption{Wavelet Coherence power spectrum (WCPS) of the  water level standardized daily anomalies relations between consecutive Gauge stations: (a) Tefe-Manaus, (b) Manaus-Óbidos. The U-shaped grid form  shows the cone of influence. The 5\% significance level against red noise model is shown as a thick black contour. The grey line into the WCPS denotes the 32-128-days scale-average power zoom  from the Morlet wavelet transform variance: (c) Tefe and Manaus, (d) Manaus and Óbidos}
     \label{fig:W3}
  \end{center}
 \end{figure}

Dealing with non-stationary processes, as in hydrology, the use of a time-frequency representation of the signals is suggested. Such coherence highlights the temporal variations of the correlation between two signals and allows the detection of transient of high covariance  \cite{Labatetal2005}. To overcome the problem inherent to non-stationary signals, it has been proposed to introduce the wavelet coherence  \cite{TorrenceCompo1998}. The wavelet coherence is a cross wavelet power and reveals areas with high common power \cite{Grinsted04}. This means how coherent the cross wavelet transform of two time series are in the time-frequency space. Regardless of the scale, wavelet coherence of water level between consecutive gauges station first reflects a global coherence within the 95\% of confidence interval, especially for the Manaus-Óbidos pairwise standardized anomalies (Figure \ref{fig:W3} a, b). 
 
The wavelet coherence between  consecutive gauges station highlights a near permanent coherence for annual processes as expected, this means that hydro-climatological cycle are invariable throughout the time.  In the Figure \ref{fig:W3} (a, b) it can be observed that the 5\% significant coherence starts next to a month period, denoting that high frequency processes on each gauge station are independent and stochastic. When examining a scale-period 28-128-days processes, Manaus-Óbidos wavelet coherence values are higher than 0.9, whereas the wavelet coherence of Tefe-Manaus water level is lower, more intermittent and higher values lies between 0.80 and 0.90. The WT variability related to this scale-period is presented in the Figure \ref{fig:W3} (c,d) for providing a better understanding of how the WT amplitude influences in the WCC throughout the time-series. 

In the Tefe-Manaus river stretch the low coherence is associated to the gauges localization (Figure \ref{fig:1}), Tefe is placed near the Solimões/Amazon river source and it is influenced by different rainfall regimes. By the other hand, the Manaus gauge station is influenced by two large rivers Negro and Solimões. In the Manaus-Óbidos river stretch both gauges are influenced by the same rivers and this feature is captured by the WT of the series and denotes a hight coherence in the WCC between these gauges. 

For decadal analysis in the  28-128-days scale-period, for the Tefe-Manaus pairs  was observed the strongest WCC  during  the 1980s decade. This might be related to the gaps into the data series and the pairwise selection of the available data. From 1990s to 2000s decades this coherence is lower in average throughout the time-series, but punctually when the WT amplitude signal were in face the coherence values become higher (Figure \ref{fig:W3} a, c). This characteristic suggest that since 1990 the  environmental factors  affecting the relationship between Tefe-Manaus consecutive gauge stations remain constant too. 

In the Manaus-Óbidos pair was observed that even with low amplitude the WT variability 32-128-days scale-period were in face almost all the 1980s and 1990s decades, so the WCC average values were high and almost permanent throughout the time-series (Figure \ref{fig:W3} b, d) and for the 2000s decade was observed a WCC average value diminution. This feature could be explained by the waves flattened because of the relative 32-128-days reduction and 64-256-days increase in the waves amplitude for the last three decades (Table \ref{tabela:serie}).  This characteristic might be associated with land cover changes in the floodplains area, for example, with the forest removal the system responds more quickly to water ingress, by diffuse or concentrated input, but the water remain in the system more time because of environmental changes that influence into the natural water flow, for example: sediment transport and deposition, erosion and others.

\section{Final considerations}

In particular for infra-annual periods was possible to detect changes for the water level variability along the Solimões/Amazonas main channel, by applying the Morlet Wavelet Transformation (WT) and  Wavelet Cross Coherence (WCC) methods. This methodology is a recent tool for hydrological analyses, in particular for the water level historical series signals have not been used before.

It was possible to quantify the waves amplitude for the WT infra-annual scaled-period and were quite similar to the three gauges station denoting that the water level variability are related to a same hydrological forcing functions. Changes in the WCC was detected for the Manaus-Óbidos river stretch and this characteristic might be associated with land cover changes in the floodplains.

This study hypothesizes that the progressive forest removal in Solimões/Amazonas floodplain in the last 30 years might be the main environmental factor affecting time variability in the correlation coefficient. The next step is to test this hypotheses by integrating land cover changes into the floodplain along the time with hydrological modelling simulations.

\section{Acknowledge}

The authors are grateful for the financial and operational support from the Brazilian agencies FAPESP (São Paulo Research Foundation, grants 2011/00250-1 and 2012/21877-5) and INPE (National Institute for Space Research). Also we would like to thank for the constructive comments provided by anonymous reviewers.

\bibliographystyle{witpress}
\bibliography{sample}

\begin{thebibliography}{10}

\bibitem{Junk1997}
Junk, W.J., \emph{The central Amazon floodplain: Ecology of a pulsing system}.
  Springer: Berlin, 1997.

\bibitem{ALCANTARAetal2010}
Alcântara, H., Novo, E., Stech, J., Lorenzzetti, J., Barbosa, C., Assireu, A.,
  \& Souza, A., A contribution to understanding the turbidity behaviour in an
  amazon floodplain. \emph{Hydrology and Earth System Sciences},
  \textbf{14(2)}, pp. 351--364, 2010.

\bibitem{Papaetal2008}
Papa, F., Güntner, A., Frappart, F., Prigent, C. \& Rossow, W.B., Variations of
  surface water extent and water storage in large river basins: A comparison of
  different global data sources. \emph{Geophysical Research Letters},
  \textbf{35(11)}, p. L11401, 2008.

\bibitem{Hessetal2003}
Hess, L., Melack, J., Novo, E.M.L.M., Barbosa, C. \& Gastil, M., Dual-season
  mapping of wetland inundation and vegetation for the central amazon basin.
  \emph{Remote Sensing of Environment}, \textbf{87}, pp. 404 -- 428, 2003.

\bibitem{Renoetal2011Viv}
Renó, V.F., Novo, E.M., Suemitsu, C., Renno, C.D. \& Silva, T., Assessment of
  deforestation in the lower amazon floodplain using historical landsat
  {MSS/TM} imagery. \emph{Remote Sensing of Environment}, \textbf{115(12)}, pp.
  3446 -- 3456, 2011.

\bibitem{Birkett1998}
Birkett, C.M., Contribution of the topex nasa radar altimeter to the global
  monitoring of large rivers and wetlands. \emph{Water Resources Research},
  \textbf{34(5)}, pp. 1223--1239, 1998.

\bibitem{ALMEIDA2012}
Almeida, F.G., Calmant, S., Seyler, F., Ramillien, G., Blitzkow, D., Matos, A.
  \& Silva, J.S., {Time-variations of equivalent water heights'from Grace
  Mission and in-situ river stages in the Amazon basin}. \emph{{Acta
  Amazonica}}, \textbf{42}, pp. 125 -- 134, 2012.

\bibitem{KazezyilmazAlhaneMedina2007}
Kazezyilmaz-Alhan, C.M. \& Medina, M.A., Kinematic and diffusion
  waves:analytical and numerical solutions to overland and channel flow.
  \emph{Journal of Hydraulic Engineering}, \textbf{133(2)}, pp. 217--228, 2007.

\bibitem{Nobreetal2006}
Nobre, P., Marengo, A., Cavalcanti, I.F.A., Obregon, G., Barros, V., Camilloni,
  I., Campos, N. \& Ferreira, G., Seasonal to decadal predictability and
  prediction of south american climate. \emph{J Climate (Special Section)},
  \textbf{19(23)}, pp. 5988--6004, 2006.

\bibitem{TorrenceCompo1998}
Torrence, C. \& Compo, G., A practical guide to wavelet analysis.
  \emph{Bulletin of the American Meteorological Society}, \textbf{79}, pp.
  61--78, 1998.

\bibitem{Grinsted04}
Grinsted, A., Moore, J.C. \& Jevrejeva, S., Crosswavelet and wavelet coherence.
  \emph{Nonlinear Processes in Geophysics}, \textbf{11}, pp. 561--566.

\bibitem{pierro09}
Pierro, D.M., \emph{Web2py}. 1st edition, p. 246.

\bibitem{Zhanetal2010}
Zhang, Q., Chong-Yu, X. \& Chen, D., Wavelet-based characterization of water
  level behaviors in the pearl river estuary, china. \emph{Stoch Environ Res
  Risk Assess}, \textbf{24}, pp. 81--92, 2010.

\bibitem{Marengoetal1998}
Marengo, J., Tomasella, J. \& Uvo, C., Long-term streamflow and rainfall
  fluctuations in tropical south america: Amazonia, eastern brazil and
  northwest peru. \emph{Journal of Geophysical Research}, \textbf{103}, pp.
  1775--1783, 1998.

\bibitem{Labatetal2005}
Labat, D., Ronchail, J. \& Guyot, J., Recent advances in wavelet analyses: Part
  2. amazon, parana, orinoco and congo discharges time scale variability.
  \emph{Journal of Hydrology}, \textbf{314}, pp. 289--311, 2005.

\bibitem{Uvoetal1998}
Uvo, C.R.B., Repelli, C.A., Zebiak, S. \& Kushnir, Y., The relationship between
  tropical pacific and atlantic sst and northeast brazil monthly precipitation.
  \emph{J Climate}, \textbf{11}, pp. 551--562, 1998.

\end{thebibliography}

\end{document}